\documentclass[%
 reprint,
 amsmath,amssymb,
 aps,
]{revtex4-2}

\newcommand{\mycomment}[1]{}
\usepackage{booktabs}
\usepackage{comment}
\newcommand{\kBT}{k_{\mathrm{B}}T}
\usepackage[most]{tcolorbox}
\definecolor{cream}{RGB}{222,217,201}
\usepackage{xcolor} 
\usepackage{enumitem}
\usepackage{graphicx}
\usepackage{dcolumn}
\usepackage{bm}
\usepackage[hidelinks]{hyperref}

\begin{document}

\preprint{APS/123-QED}

\title{A tutorial for mesoscale computer simulations of lipid membranes: \\tether pulling, tubulation and fluctuations}

\author{Maitane Muñoz-Basagoiti}
\email{These authors contributed equally.}
\author{Felix Frey$^*$}
\author{Billie Meadowcroft$^*$}
\author{Miguel Amaral$^*$}
\author{Adam Prada$^*$}
\author{Andela Šarić}%
\email{andela.saric@ista.ac.at}
\affiliation{%
 Institute of Science and Technology Austria, Am Campus, 3400 Klosterneuburg (Austria)
 }%

\date{\today}

\begin{abstract}
Lipid membranes and membrane deformations are a long-standing area of research in soft matter and biophysics.
Computer simulations have complemented analytical and experimental approaches as one of the pillars in the field.
However, setting up and using membrane simulations can come with barriers due to the multidisciplinary effort involved and the vast choice of existing simulations models.
In this review, we introduce the non-expert reader to coarse-grained membrane simulations (CGMS) at the mesoscale.
Firstly, we give a concise overview of the modelling approaches to study fluid membranes, together with guidance to more specialized references.
Secondly, we provide a conceptual guide on how to develop CGMS.
Lastly, we construct a hands-on tutorial on how to apply CGMS, by providing a pedagogical examination of tether pulling, tubulation and fluctuations with
three different membrane models, and discussing them in terms of their scope and how resource-intensive they are.
To ease the reader's venture into the field, we provide a repository with ready-to-run tutorials.

\end{abstract}

\maketitle

\section{Introduction}
One of the most astonishing results of biological evolution is the genesis of fluid lipid membranes, which have evolved to be remarkably versatile and ubiquitous in biological systems.
The relevance of this evolutionary step can be best seen in the outer plasma membrane and the compartmentalization of biological cells, which are fundamentally achieved via fluid lipid membranes. This includes the membrane-enclosed organelles such as the endoplasmic reticulum, the Golgi apparatus, mitochondria or the cell nucleus~\cite{alberts2015}.
One could naïvely assume that in order to shield the interior of a compartment from the exterior, membranes will be rigid and stable, but in fact the opposite is true\cite{bassereau2018}.
Cellular membranes exhibit a rich phenomenology of shapes that are drastically remodelled across different length scales and in various cellular processes such as endo- and exocytosis, cell adhesion, cell migration and cell division~\cite{frey2021}.
This shows that fluid lipid membranes combine complementary physical properties:
on the one hand, they spontaneously self-assemble from lipid molecules in an aqueous environment, forming closed surfaces that define and separate the inner from the outer;
on the other hand, due to the relatively weak forces acting between the lipids that form them, lipid membranes typically operate in the fluid phase in the cell environment and are thus an extremely flexible and malleable material~\cite{phillips2012}.


\begin{figure*}[t] 
    \centering

    \begin{tcolorbox}[
        width=\textwidth,          
        height=13.5cm,                
        colframe=white,             
        colback=cream,             
        arc=5pt,                  
        boxrule=1mm,               
        left=10pt, right=10pt,     
        top=10pt, bottom=10pt,
        valign=center              
    ]

    \begin{minipage}[c][8cm][c]{0.3\textwidth}
    \centering
    \includegraphics[width=\textwidth]{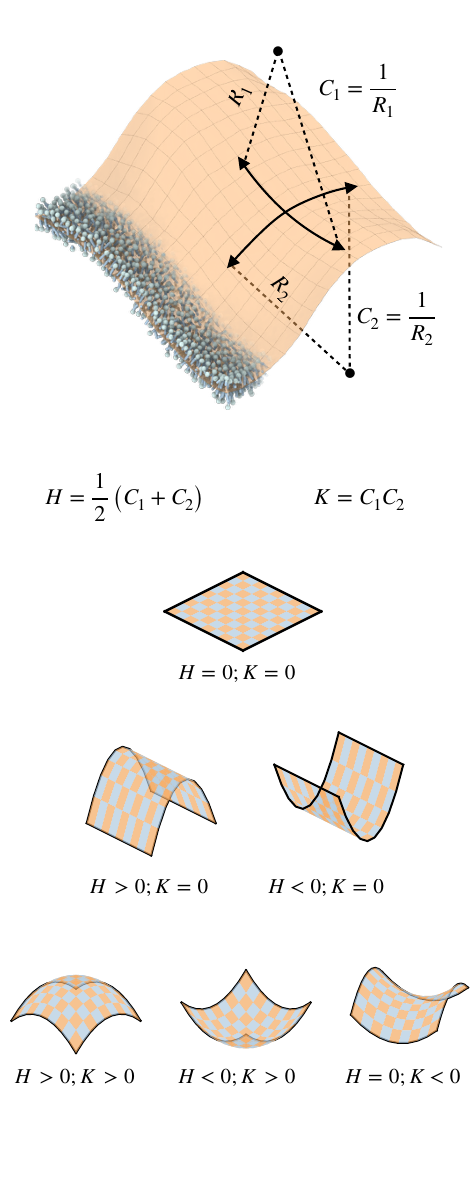}
    \end{minipage}
    \begin{minipage}[c][8cm][c]{0.65\textwidth}
    \renewcommand\labelitemi{\tiny$\bullet$}
    \textbf{BOX I: Basics of membrane physics}
    \begin{itemize}
    \item \textbf{Geometry of surfaces:}
    A lipid membrane can be described as a 2D surface embedded in a 3D space. To describe its shape, it is necessary to define the two \textit{principal curvatures}, $C_1$ and $C_2$, at any point of the surface. The magnitude of $|C_i|=1/R_i$ and $C_{i}>0$ if the surface bulges outwards and vice versa. Two combinations of the principal curvatures known as the \textit{mean} curvature $H=(C_1+C_2)/2$ and \textit{Gaussian} curvature $K=C_1 C_2$ are used to describe the geometry of a surface and classify its type: planes and cylinders have zero Gaussian curvature ($K=0$), spheres have positive Gaussian curvature ($K>0$) and saddles have negative Gaussian curvature ($K<0$); similar to the plane, saddle shapes can also have zero mean curvature ($H=0$).

    \item \textbf{Energetics of membranes as surfaces:}
    The \textit{Helfrich-Hamiltonian} $\mathcal{H}$ is a function, or precisely a surface based energy functional~\cite{phillips2012}, which assigns an energy density to every point of the surface in a lipid membrane based on the values of $H$ and $K$; it results in the \textit{membrane shape energy} when integrated along the surface. At equilibrium, a lipid membrane adopts a shape that minimizes $\mathcal{H}$.
    To find such a minimum, variational calculus, the Euler-Lagrange equations or various numerical methods can be applied.
    This minimization process can also be subject to additional constraints by using Lagrange multipliers. For example, one can also choose to minimise the surface area by introducing membrane tension $\gamma$ as a constraint.

    \item \textbf{Triangulated meshes:}
    A surface can be represented using a discrete collection of vertices, edges and faces referred to as a \textit{mesh}. The vertices of a mesh are points in 3D space which are connected by edges, which then delimit the faces. In a \textit{triangulated} mesh, the faces of the mesh are (not-necessarily equilateral) triangles, and each face has an associated perpendicular unit vector which defines the orientation of the corresponding triangle. The relative angle between the normal vectors of adjacent faces is given by their scalar product and it can be used to describe the local curvature of the surface~\cite{siggel2022}.
    \end{itemize}

        \end{minipage}
    \end{tcolorbox}
\end{figure*}

Due to their biological relevance, fluid lipid membranes have been studied for decades using experimental, theoretical and computational approaches.
As a result, the field of membrane biophysics has experienced many breakthroughs and it is relatively mature compared to other areas of biophysics. It has been extensively reviewed from the experimental~\cite{dimova2019,idema2019,turlier2019,Bassereau2014,mcmahon2005}, theoretical~\cite{lipowsky2021,frey2021,deserno2015,seifert1997,lipowsky1991} as well as computational perspectives~\cite{beiter2024,li_mesoscopic2024,dasanna2024,kumar2022}.
Nonetheless, the field is far from being saturated.
With the advent of new imaging concepts such as cryo-electron microscopy and super-resolution microscopy, the experimental sciences have recently experienced a resolution revolution \cite{kuhlbrandt2014,sahl2017}. Biological phenomena that take place on membranes can now be visualized with nanometer resolution, and investigation of previously unobservable processes is becoming tractable.
For example, it is now possible to understand with an unprecedented resolution how supra-molecular protein assemblies such as the dynamic ESCRT-III filaments and the clathrin coats remodel membranes~\cite{pfitzner2020,mund2023}, or how changing the membrane composition can lead to membrane fission~\cite{steinkuhler2020}.
Likewise, in recent years the material properties of fluid lipid membranes have gathered the interest of the soft matter community, with membranes being used as platforms for self-assembly~\cite{azadbakht_wrapping_2023, azadbakht_nonadditivity_2024} and biomimetic design~\cite{willems_phase_2023}.

The study of biological problems often requires the creation of sophisticated models -- ideally ones that are easy to use, modify and extend. Fortunately, technological and scientific progress allows us to perform ever more complex and resource-intensive calculations. Therefore, computer simulations have found a wide range of applications including the study of membrane pores \cite{marrink2009}, membrane trafficking~\cite{alvarez2023}, membrane remodelling~\cite{kumar2022,larsen2022}, membrane fission and fusion~\cite{noguchi2009}, membrane deformation in flow~\cite{guckenberger2017,dasanna2024} and membrane organelle shapes~\cite{pezeshkian2021}.

State-of-the-art modelling of the open problems in membrane biophysics requires a variety of expertise, ranging from cell biology and membrane biophysics to programming.
It can be challenging for beginners and non-experts to start developing and applying membrane models due to the multidisciplinary character of the questions, the long-standing tradition of the field, and the wealth of knowledge and models.
However, choosing a model should be a conscious decision which is, ideally, not influenced by mere technical hurdles or lack of expertise.
Before one chooses an appropriate membrane model it is therefore indispensable to know about the limitations of the different options and common pitfalls.
The ambition of this review is to \textit{guide} the reader in their choice of models when they want to start simulating fluid lipid membranes.
This review is specifically targeted to a broad audience of researchers that are interested either in starting with lipid membrane simulations, such as computational physicists, or in getting a better understanding of the computational methods, such as experimental biologists with less experience in computer simulations.
We limit the scope of the review to problems that involve mesoscale membrane deformations for which coarse-grained modelling approaches are most suitable.

We aim to accomplish three tasks.
First, for the sake of pedagogical completeness, we give a concise overview of fluid lipid membrane models and corresponding simulation techniques.
Second, we present a conceptual guide on how to develop coarse-grained membrane simulations.
Third and most importantly, we provide a hands-on tutorial on how to apply mesoscale membrane simulations to three classical membrane problems: force required to extrude a membrane tube, equilibrium radius of the extruded tube, and membrane fluctuations.
Our focus is on a comparison of different models and a pedagogical examination of them in terms of what these models can achieve and how resource-intensive they are.
A repository with ready-to-run tutorials to simulate three different models is provided alongside the review~\cite{githubrepo}.
Whenever possible, we point the reader to other relevant reviews which expand on specific topics and technical details.

\section{Overview of membrane models}
In order to describe fluid lipid membranes, two main approaches have been developed over the last five decades:
Membranes are either represented based on \textit{continuous surfaces} (\ref{section:surface}) or based on \textit{discrete building blocks} (\ref{section:particle}).
In the following sections we briefly outline the two modelling approaches.

\begin{figure*}[th!]
    \centering
    \includegraphics[width=0.92\textwidth]{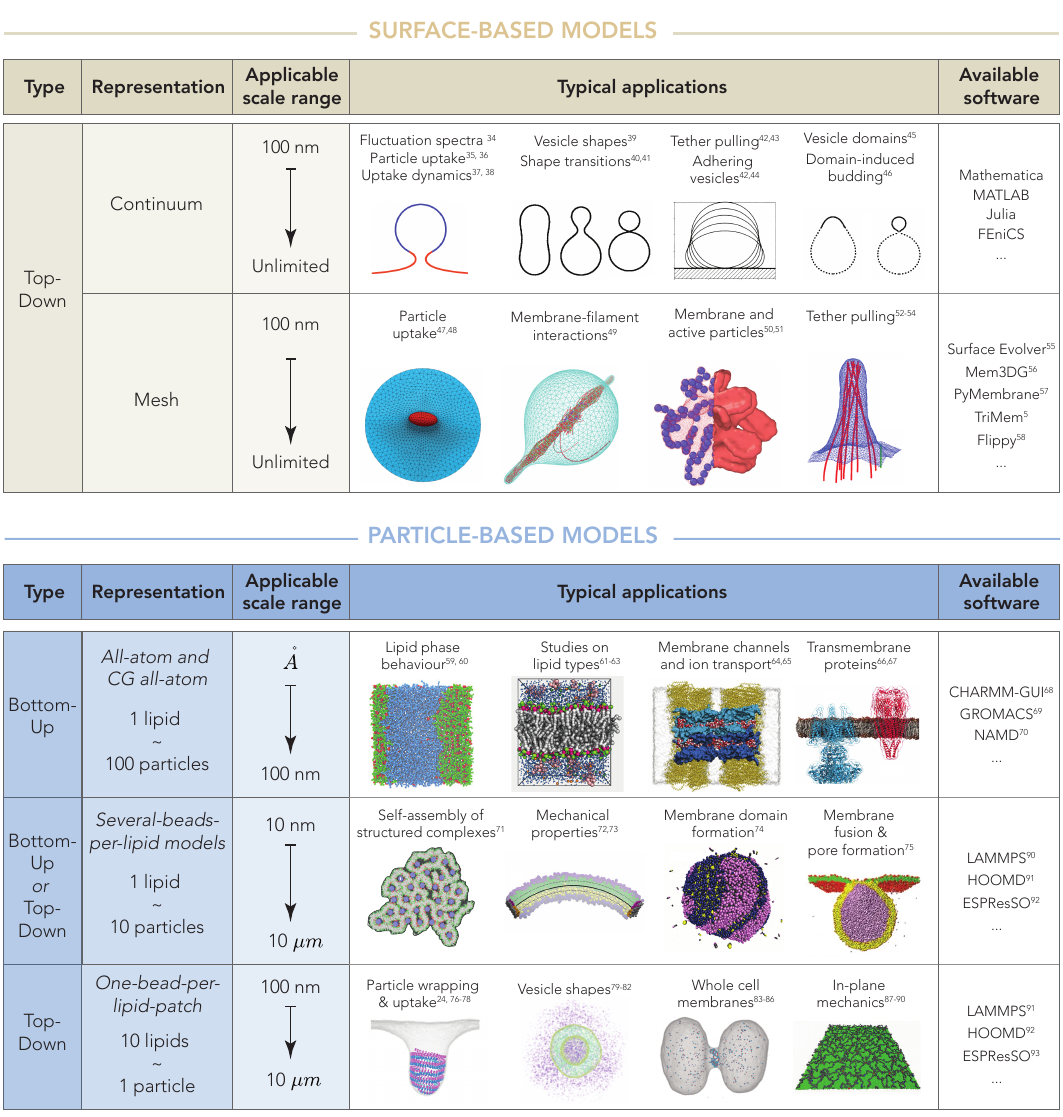}
    \caption{\textbf{Classification and applications of fluid membrane models.} Continuum models have been used to study membrane fluctuation spectra~\cite{Helfrich1984}, particle uptake~\cite{Deserno2004,Foret2014}, uptake dynamics~\cite{Frey2019,frey2019b},  vesicle shapes~\cite{Seifert1991}, shape transitions~\cite{Christ2021,Frey2022}, tether pulling~\cite{Smith2004, Derenyi2002}, adhering vesicles~\cite{Seifert1990,Smith2004}, vesicle domains~\cite{Jülicher1996},and domain induced budding~\cite{Jülicher1993}among others. Applications of mesh models are particle uptake~\cite{dasgupta2013,dasgupta2014}, membrane-filament interactions~\cite{ni2021}, membranes and active particles~\cite{vutukuri2020,iyer2023} and tether pulling~\cite{pezeshkian2024, weichsel_more_2016, koster_force_2005}. Available mesh model software are the Surface Evolver~\cite{Brakke1992}, Mem3DG~\cite{zhu2022}, PyMembrane~\cite{matoz-fernandez_pymembrane_2023}, TriMem~\cite{siggel2022} or Flippy~\cite{dadunashvili_flippy_2023} .Particle based models representing 1 lipid as a collection of 10-100 particles have been applied to study lipid phase behavior~\cite{Carpenter2018,Davis2013}, studies on lipid types~\cite{Wassenaar2015,Jo2009,Qi2015}, membrane channels and ion transport~\cite{Yesylevskyy2010,KhaliliAraghi2009} and transmembrane proteins~\cite{Mandal2021,Muller2019}. These models can be simulated using software like CHARMM-GUI~\cite{CHARMGUI}, GROMACS~\cite{Gromacs} or NAMD~\cite{NAMD}. Simulations with lipids represented as 1-10 particles (which we refer to as \textit{several-beads-per-lipid} models in section~\ref{subsection:severalbeadsperlipid}), and focused on the 10nm - 10 micron scale have been applied to studying the self-assembly of lipids and other biomolecules into structured complexes~\cite{farago_simulation_2009}, mechanical properties~\cite{foleyAsymmetricMembraneSticky2024,huDeterminingGaussianCurvature2012}, membrane domain formation~\cite{cooke2005b} and membrane fusion and pore formation~\cite{Shillock2006}. Simulation models where a patch of membrane is represented as a single particle (\textit{one-bead-per-lipid-patch} models in section~\ref{section:onebeadperlipid}) have been exploited to probe phenomena at larger scales (100 nm-10 microns) such as particle wrapping and uptake~\cite{jiang_modelling_2022,bao_experimental_2021,forster_exploring_2020,azadbakht_wrapping_2023}, vesicle shapes~\cite{Rower2019HeterogeneousVW,campos_dynamics_2020,huang_coupled_2011,yuan_dynamic_2010}, whole cell membranes~\cite{harker-kirschneck_physical_2022,singh_bactericidal_2022,becton_flow_2019,yu_temperature-_2019} and in-plane mechanics~\cite{ahmadpoor_entropic_2022,debets_characterising_2020,zhang_multiple_2015,sadeghi_thermodynamics_2021}, among others. Available software to simulate these types of models are LAMMPS~\cite{Thompson2022}, HOOMD~\cite{anderson_hoomd-blue_2020} or ESPResSO~\cite{weik_espresso_2019}.}
    \label{tab:Table1}
\end{figure*}
\newpage
\clearpage

\subsection{Surface-based models}\label{section:surface}
Surface-based models neglect the thickness of the membrane, and instead describe the membrane as a 2D surface in a 3D space.
The surface is represented either by a continuous function (\ref{subsubsec_continuummodels}) or by a discrete mesh approximating a continuous surface (\ref{subsection_meshes}). While analytic continuum models have been successfully used to derive many classical results in membrane biophysics, in general, when the surface cannot be parametrised by any basic geometry such as a (nearly) flat plane, a cylinder, a sphere, or combinations of those, it is often easier to represent the surface by a triangulated mesh to be able to carry out the calculations.


\subsubsection{Continuum models}\label{subsubsec_continuummodels}
Continuum models for fluid lipid membranes have been the subject of several excellent reviews~\cite{seifert1997,deserno2015,lipowsky2021}.
In this section we only introduce the general idea behind the models and discuss several classical problems of membrane biophysics which we later use as a reference for comparison with our computer simulations.

The fundamental assumption of continuum models is that fluid lipid membranes, which are only 4--5~nm thick but can span over micrometer scales, can be represented by a 2D surface without thickness in a 3D space.
Due to the assumed separation of length scales (difference between thickness and size), the validity range of such models is unlimited at the upper end, and it is about 100~nm at the lower end (cf.~Fig.~\ref{tab:Table1}).
If we are able to express the energy of the membrane, we can find its equilibrium shape by minimising this energy.
The membrane shape energy \textit{functional} $\mathcal{H}$ is based on the observation that lipid membranes are \textit{incompressible fluids}, i.e. it is hard to stretch or compress a membrane within the membrane plane, and due to the high diffusivity of lipids within this plane, the membrane  is fluid (i.e., it has a vanishing shear modulus).
As a consequence, the only relevant energetic change when deforming a fluid lipid membrane is the out-of-plane bending of the surface, which is usually quantified through membrane curvatures (cf. Box I). The membrane shape energy $\mathcal{H}$ is then defined as a function of the mean curvature $H$ and the Gaussian curvature $K$.
The most commonly used membrane Hamiltonian (energy functional) is the so-called Helfrich Hamiltonian (or Helfrich-Canham-Evans Hamiltonian)~\cite{helfrich1973}, which expands the energy up to the leading order in $H$ and $K$.
This is the second order for $H$ and first order for $K$ as can be seen from the equation of this Hamiltonian ($K$ carries the units of 1 over length squared similar to $H^2$)
\begin{equation}
\mathcal{H}=2 \kappa \int (H-H_0)^2 \mathrm{d} A + \bar{\kappa} \int K \mathrm{d} A +
\gamma \int \mathrm{d} A \, .
\label{eq:Helfrich}
\end{equation}
In Eq.~(\ref{eq:Helfrich}), $\kappa$ quantifies the membrane's resistance to bending and $\bar{\kappa}$ the resistance to changes in the membrane topology, e.g. during membrane fusion or fission.
Typically, for fluid membranes the membrane bending rigidity is in the range of $\kappa \in [10, 100] ~\kBT$, which can be measured by pulling membrane tethers or analysing membrane fluctuations ~\cite{Bassereau2014}.
By contrast, the Gaussian curvature modulus is hard to measure since any attempt requires that the membrane changes its topology.
Using continuum arguments and coarse-grained simulations, the Gaussian curvature modulus is expected in the range of~\cite{deserno2015,huDeterminingGaussianCurvature2012} $\bar{\kappa}\in [-0.5, -1]\kappa$.
The quantity $H_0$ is the \textit{preferred} or \textit{spontaneous} membrane curvature, and it describes the membrane's tendency to curve spontaneously, due to, e.g. an asymmetric lipid composition or proteins.
The last term in Eq.~(\ref{eq:Helfrich}) imposes a constraint on the area of the membrane, where changing of the membrane area comes with an energetic cost given by the membrane tension $\gamma$.
From the physical point of view, $\gamma$ acts as a Lagrange multiplier or the chemical potential for membrane area~\cite{farago2003_thermal,deserno2015} and it is of the order of~\cite{rangamani2022} $\gamma \in [1\times10^{-6}-5 \times 10^{-4}]\, \mathrm{N/m}$.
The membrane shape energy can be computed by integrating all terms in Eq.~(\ref{eq:Helfrich}) over the surface area of the membrane.
Eq.~(\ref{eq:Helfrich}) is the simplest representation of a membrane shape energy. However, more elaborate membrane energy functionals have been developed in order to represent, for instance, the asymmetry between the two phospholipid leaflets that make up most membranes due to effects such as asymmetric lipid compositions, membrane proteins or preferential protein binding.
Such effects can be taken into account by adding a non-local curvature energy, like the area-difference-elasticity~\cite{seifert1997}.

The Helfrich Hamiltonian has been successfully used to study several classical problems of membrane biophysics (see first row in Table~\ref{tab:Table1}).
The equilibrium shapes of membrane vesicles can be calculated by finding the membrane configuration that has the smallest energy~\cite{Seifert1991}.
In this specific case it means that by minimizing the energy functional under the constraint of enclosed volume, the corresponding Euler-Lagrange equation --- typically called the \textit{shape equation} --- predicts the equilibrium membrane shape~\cite{zhong-can1989}.
In general the shape equation is a non-linear Partial Differential Equation (PDE) that cannot be easily solved.
However, by assuming rotational symmetry of the membrane shape it is possible to remove some of the complexity of the problem and simplify the PDE to a system of non-linear Ordinary Differential Equations (ODEs).
Such systems of ODEs have been solved numerically to predict the shape diagram of lipid membrane vesicles~\cite{Seifert1991} (cf.~Fig.~\ref{tab:Table1}).
Other similar uses of continuum models include the study of vesicle adhesion~\cite{Seifert1990,Smith2004}, of shape transitions of vesicles and functionalised vesicles~\cite{Christ2021,Frey2022}, of domain induced budding~\cite{Jülicher1993} and of vesicle domains~\cite{Jülicher1996} on the micrometer scale (cf.~Fig.~\ref{tab:Table1}).

Aside from predicting equilibrium vesicle shapes, continuum theory can determine the response of membranes in standard experimental assays typically used to characterize the membrane properties -- see reference \citenum{Bassereau2014} for a review on testing continuum membrane models experimentally. For example, continuum theory connects the properties of a membrane to its height fluctuation spectrum, relating the amplitude of the fluctuations ($h$) in the Fourier space to the bending rigidity and the membrane tension as $h(q)^2 \propto \kBT/(\kappa q^4+\gamma q^2)$, where $q$ is the wave number~\cite{helfrich_undulations_1984}.
It also predicts the force $f=2\pi \sqrt{2\gamma \kappa}$ required to pull a membrane tube and its corresponding equilibrium radius $R_\mathrm{eq}=\sqrt{\kappa/(2\gamma)}$~\cite{Derenyi2002}.
In addition, continuum theory has also been used to study particle wrapping by membranes, e.g. for predicting the shape of the membrane during particle uptake~\cite{Deserno2004,Foret2014} and its dynamics~\cite{Frey2019,frey2019b} (cf.~Fig.~\ref{tab:Table1}), or for determining of the regions in phase space in which membrane adhesive particles are completely, partially or not at all wrapped as a function of bending stiffness, adhesion energy and membrane tension~\cite{Deserno2004}.
Membrane particle uptake has been reviewed in detail previously~\cite{Bahrami2014}.

\begin{figure*}[t] 
    \begin{tcolorbox}[
        width=\textwidth,          
        height=6.5cm,                
        colframe=white,             
        colback=cream,             
        arc=5pt,                  
        boxrule=1mm,               
        left=10pt, right=10pt,     
        top=10pt, bottom=10pt,      
    ]

\textbf{BOX II: Sampling configurations using Molecular Dynamics and Monte Carlo simulations}

\begin{enumerate}[leftmargin=1mm]
    \item [] Molecular Dynamics (MD) is a simulation method is based on the numerical integration of Newton's equations of motion for a group of particles to obtain their trajectories in real time. These trajectories can then be used to read off static (characterising the state and properties of the system) such as the temperature and pressure, or dynamic observables (characterising the trajectories) like the diffusivity and time correlations of particles. The time evolution of the equations of motion relies on the calculation of the forces acting on all particles in the system for the given timestep, which depend on the interaction potentials between particles and the specific particle positions. Unlike the MD simulations, the Monte Carlo (MC) simulations do not rely on the propagation of equations of motion but are instead based on efficiently sampling the equilibrium distribution that characterises the system. This means that the evolution of the system in the simulation time does not necessarily bear relevance to the physical motion of the system. Although they can be used to obtain dynamical properties in certain limits, they are typically used to generate equilibrium configurations. An excellent introduction to MD and MC simulations can be found in~\cite{understandingsims_2023}. Established software to conduct MD and MC simulations include LAMMPS~\cite{Thompson2022} (see~\cite{lammpstutorials} for introductory tutorials), HOOMD~\cite{anderson_hoomd-blue_2020} or ESPRESSO~\cite{weik_espresso_2019}.
\end{enumerate}
\end{tcolorbox}
\end{figure*}

\subsubsection{Mesh models}\label{subsection_meshes}

Analytic solution of continuum membrane models is mostly limited to cases that show rotational symmetry.
In more complex systems, as when the membrane is coupled to discrete proteins or cytoskeletal filaments, or when it interacts with extracellular structures, PDEs describing the membrane shape can be obtained, but these are difficult to solve without further assumptions.
An alternative approach to a straightforward PDE solution is to determine the membrane shape by generating configurations that gradually minimise the energy described by Eq.~(\ref{eq:Helfrich}). A particularly powerful approach to carry out such an energy minimisation is via the representation of the surface as a triangulated network or \textit{mesh} of vertices connected by edges (cf. Box I). The membrane shapes that minimise the surface energy can then be determined using Monte-Carlo (MC) or Molecular Dynamics (MD) simulations (cf. Box II), as well as other numerical techniques like using the Surface Evolver~\cite{Brakke1992,dasgupta2014}.
Below we give only a brief introduction to triangulated mesh models for lipid membrane simulation, and we direct the reader to in-depth reviews for further details~\cite{gompper1997,kumar2022}. A hands-on application of triangulated meshes for lipid fluid membrane simulation is then provided in section~\ref{subsec:trilmp}.

In a triangulated network, bending energy can be accounted for in different ways. To simulate \textit{tethered} membranes, which are characterized by a fixed network connectivity, bending costs can be implemented by combining bond and dihedral angle potentials in order to keep the area of each triangle in the mesh approximately constant. Nonetheless, to simulate \textit{fluid} membranes a more sophisticated discretization of the bending energy introduced in Eq.~(\ref{eq:Helfrich}) is needed, as the connectivity of the network must be regularly updated in order to achieve their characteristic in-plane fluidity. Such discretization can be approached in different ways. Frequently used schemes are based on the angle formed between normal vectors of adjacent triangles in the mesh (cf. Box I)~\cite{kantor_phase_1987, julicher_morphology_1996}, although alternative discretisations of the bending energy that include surface-related corrections are also available~\cite{gompper_random_1996,bian_bending_2020}. On top of the bending energy, the length of the edges in the network must be also be constrained to prevent the mesh from self-intersecting~\cite{gompper1997} (see section~\ref{subsec:trilmp}), and the ratio between the rates of edge connectivity flips and vertex displacements determines the membrane viscosity~\cite{dasanna2024}. The minimal modelling of a triangulated fluid membrane is completed by incorporating area and volume constraints into the simulation~\cite{bahrami_formation_2017, siggel2022}. An increase in membrane area can be penalized through a discretized version of the linear constraint in Eq.~(\ref{eq:Helfrich}) and a surface tension $\gamma>0$, where the total membrane area is equal to the sum of the area of each triangle in the network. Alternatively, it is also possible to prescribe a desired membrane area $A_0$, and penalize deviations from such value using a quadratic constraint like
\begin{equation}
    E_A = k_A \left( \frac{A - A_0}{A_0}\right)^2,
\end{equation}
where $A$ is the actual membrane area, and $k_A$ is a constant that enforces the constraint. A similar quadratic constraint can be used to prescribe a desired volume $V_0$ in simulations of closed surfaces such as lipid vesicles. Constraining the membrane area and enclosed volume opens the door to controlling the surface-to-volume ratio of simulated lipid vesicles~\cite{bahrami_formation_2017}.

Membrane shape and dynamics using triangulated membrane models can be explored with a variety of simulation techniques such as Monte Carlo (MC) or Hybrid Monte Carlo (HMC). In pure MC simulations, both the positions of the vertices in the mesh and the connectivity of the mesh are updated through trial moves. If those moves increase the energy of the system, they are rejected with some probability, which can make the sampling of the configuration space slow. HMC simulations combine MC bond-swap moves to update mesh connectivity with Molecular Dynamics (MD) time evolution to compute the dynamics of the mesh vertices. This speeds up the process of finding configurations that minimize the energy of the membrane, as long as the equations of motion are correctly integrated~\cite{bian_bending_2020, veerapaneni_numerical_2009}.

Triangulated membrane models have been successfully applied to study membrane wrapping of ellipsoidal~\cite{dasgupta2013} and non-spherical particles~\cite{dasgupta2014}, tether pulling~\cite{paraschiv_influence_2021, pezeshkian2024}, membrane-filament interactions~\cite{ni2021}, self-assembly of colloids on fluid surfaces~\cite{Saric2012} and interactions with active colloids \cite{vutukuri2020,iyer2023} (cf. second row in~Fig.~\ref{tab:Table1}).
However, these models face conceptual challenges when topological changes of the membrane shape become relevant, since it is not clear how to consistently and physically meaningfully cut and reconnect the mesh~\cite{tachikawa_golgi_2017}.
In contrast, volume control is easy to achieve, since the mesh defines a clear surface, and therefore, the enclosed volume can be easily calculated.

Numerous software packages implementing triangulated membrane models are currently available, such as Trimem~\cite{siggel2022}, Mem3DG \cite{zhu2022}, PyMembrane~\cite{matoz-fernandez_pymembrane_2023} and Flippy~\cite{dadunashvili_flippy_2023}.
TriMem will be tested in the Section~\ref{sec:tests} via TriLMP~\cite{TriLMP}, which is an in-house code that couples TriMem to LAMMPS~\cite{Thompson2022}.

\subsection{Particle-based models}\label{section:particle}
Particle-based models follow a conceptually different approach to that of continuum and surface based models. Instead of describing the membrane as a surface, the constituents of the membrane are represented as particles, and the mechanical properties of lipid membranes are integrated into the system through inter-particle interaction potentials. In these models, a \textit{particle} or \textit{bead} can represent an atom, a collection of atoms~\cite{marrink       2004, DryMartini}, a section of a lipid molecule~\cite{Cooke2005}, or a section of the membrane surface ~\cite{Yuan2010}.

It is important to note that not only the resolution but also the underlying modelling philosophy varies among the different particle-based models. Generally \textit{bottom-up} models retain chemical specificity of the lipid molecules and associated macromolecules. In contrast, in generic \textit{top-down} models~\cite{Cooke2005,Yuan2010} the particle properties and inter-particle interactions capture the emergent physical properties of the membrane, but do not automatically retain \textit{chemical} details. Therefore, the spirit of the
the latter models is similar to the surface-based models of the previous section. In what follows we describe the representative cases of each flavour of models.

\textbf{Bottom-up particle models.} The most straightforward class of model to understand in terms of modelling philosophy within this category are \textit{all-atom} models, where each atom in a lipid molecule is described as a particle. Coarse-grained bottom-up models group individual atoms into single particles, while still retaining chemical specificity. A common approach is to group $\sim 4-10$ atoms together into a single particle, as done in e.g. ~\textit{Martini} models~\cite{marrink2004, DryMartini} and other similarly coarse-grained representations~\cite{beiter2024,Izvekov2005,Saunders2013, Yesylevskyy2010}. Parameters in these models can be a derived from atomistic simulations, for instance in a systematic manner that  preserves statistical mechanical properties of lipid bilayers~\cite{beiter2024}, but can also include empirical information derived from experiment, formally mixing bottom-up and top-down approach. Chemically-specific lipid models have been relevant for instance for the study of lipid phase behaviour~\cite{Carpenter2018}, the impact of membrane proteins on the membrane~\cite{Mandal2021, Muller2019}, and the interaction of membrane channels and ion pumps with the membrane~\cite{Yesylevskyy2010, KhaliliAraghi2009}(cf. Table~\ref{tab:Table1}). It is important to note that while such models capture processes at the nanoscale, mesoscale and bulk membrane properties might not be always correctly captured~\cite{loose_changing_2024}. Details on all-atom and Martini-based models can be found in Ref.~\citenum{moradi_shedding_2019}, an overview of the field in Ref.~\citenum{VENTUROLI20061}, and a pedagogical review of chemically-resolved coarse-grained models can be found Ref.~\citenum{smith_simulation_2019}. Timescales and lengthscales in bottom-up models typically do not capture mesoscale membrane deformations\cite{beiter2024}, and are beyond the scope of this review.

\textbf{Top-down particle models.} These models typically contain a much smaller number of particles than those developed via bottom-up approaches. In \textit{several-beads-per-lipid} models (section~\ref{subsection:severalbeadsperlipid}) individual lipids are described as a collection of $\sim 3-7$ particles~\cite{Deserno_SimReview}. In a more coarse manner, in \textit{one-bead-per-lipid-patch models} a small patch of membrane can be modelled as a single particle as in the ~\ref{section:onebeadperlipid})~\cite{Yuan2010}. In these models the interactions between particles are not derived bottom-up, but are typically chosen empirically such that mesoscale membrane properties (e.g. fluidity, deformability) are correctly captured and match experimentally observed behaviour. Particle-based models have been reviewed extensively before~\cite{noguchi2009,pezeshkian2021,larsen2022}, and despite partially sacrificing lipid-level resolution they have proved very valuable for exploration of mesoscale membrane deformations. Therefore, they will be the focus of this tutorial review.

\subsubsection{\textit{Several-beads-per-lipid} models}\label{subsection:severalbeadsperlipid}
Here we introduce models that represent each lipid as several (usually 3-7) particles.
A detailed review on such models can be found in~\cite{Deserno_SimReview}.
\textit{Several-beads-per-lipid} models are generic, \textit{top-down} models that do not address specific detail of the lipids.
In these models, a lipid is represented by one hydrophilic \textit{head} particle connected to a chain of hydrophobic \textit{tail} beads, mimicking the structure of many common lipids.
The main conceptual idea behind these models is to reproduce key emergent membrane properties, such as the self-assembly of single lipids into a fluid membrane as a result of hydrophobic interactions with the solvent, and a bending rigidity of a few to a few tens of $k_\mathrm{B} T$.
For this purpose, two main approaches have been developed, which can be broadly classified as explicit or implicit solvent models.

The main player of explicit models is Dissipative Particle Dynamics (DPD)~\cite{Rudiger1999,Shillcock2002,Laradji2004}.
DPD-based membrane models explicitly consider solvent-like particles in addition to lipid particles, where DPD solvent particles are coarse-grained representations of a volume of fluid.
While explicitly simulating a solvent can incur a large computational cost,
DPD models mitigate such cost by using soft potentials to account for interaction between pairs of particles in the simulation.
In contrast to hard potentials, which diverge when particles get close, soft potentials remain finite at vanishing inter-particle distances.
This allows for larger time steps during the simulation.
However, DPD thermostats can be challenging to handle, which often limits the step-size advantage of DPD~\cite{farago2016}.
An alternative approach to DPD models is known as the \textit{phantom solvent} model~\cite{lenz_simple_2005}, in which phantom solvent particles are introduced to generate the hydrophobic interactions between lipids and solvent.
To speed-up the simulations, however, solvent particles do not interact which each other, since these interactions are computationally the most numerous and hence the most costly.

The second approach are implicit solvent models.
In this case, instead of having explicit solvent particles, special interactions are introduced that effectively mimic the hydrophobic effect.
The first model of that kind uses multi-body density-dependent attraction~\cite{drouffe_computer_1991} and models a lipid as a sphere with three parts.
The idea of a multi-body density-dependent attraction was later refined~\cite{noguchi2001}
to penalize lipids with few neighbours because these lipids are exposed to the (implicit) solvent.
In this model, a lipid consists of three beads.
Additional models have been developed, using attractions between lipid beads to mimic hydrophobicity with an implicit solvent \cite{Farago2003_water,Brannigan2004,Brannigan2005,Cooke2005,cooke2005b,wang2005,Revalee2008}.
Importantly, some of these models apply
Lennard-Jones potentials, but with an extended range, to mimic the hydrophobic interaction \cite{Farago2003_water,Brannigan2005,Cooke2005,cooke2005b}.
By tuning model parameters like the bond angle stiffness~\cite{Brannigan2005} or the lipid-lipid interaction strength~\cite{Cooke2005}, it is possible to control the self-assembly of lipids into lipid bilayers, and to control membrane properties such as the bending stiffness and fluidity.

\textit{Several-beads-per-lipid} models have enough detail to allow for the simulation of lipids with variable interactions, which has lead to studies on lipid domain formation and phase separation~\cite{cooke2005b,Laradji2004}. As the membrane leaflet thickness is explicitly modelled, these models are also suitable to study the physics of membrane fusion and pore formation~\cite{Shillock2006,Zun2005}.
Additionally, given their highly coarse-grained nature, several-beads-per-lipid models enable the simulation of membranes on longer timescales ($\sim 10^{-1} ms$) and larger lengthscales ($\sim 1\mu \mathrm{m}^{2}$) than those available to all-atom or Martini-level models, making them an attractive choice for biologically measurable events which involve large-scale deformations such as particle uptake~\cite{vachaReceptorMediatedEndocytosisNanoparticles2011}.
Lastly, the combination of the computational efficiency and lipid details allows several-beads-per-lipid models to be used to interrogate mechanical properties of membranes such as the bending and Gaussian curvature moduli, and bilayer asymmetry~\cite{foleyAsymmetricMembraneSticky2024,huDeterminingGaussianCurvature2012,denOtter2003}.
Nonetheless, for the simulation of larger length- and longer timescales, like experiments with GUVs (Giant Unilamellar Vesicles) or whole cells, more coarse-grained models are required.

\subsubsection{\textit{One-bead-per-lipid-patch} models}\label{section:onebeadperlipid}

To simulate membranes at the largest possible scale with particle-based models, a high level of coarse-graining has been achieved by representing a patch of lipids by just one bead.
Several similar one-bead models deploying anisotropic potentials have been developed \cite{noguchi_meshless_2006,ballone_simple_2006,kohyama2009,Yuan2010}.
One-bead models are unique in that they have a low computational cost due to extensive coarse-graining and, as beads are not connected (i.e. models are \textit{meshless}), they can easily undergo topological changes such as the neck fission during endocytosis.
These model features have led to simulations over a large membrane area with large deformations, and to modelling phenomena where membranes are cut (cf. Fig.~\ref{tab:Table1}).
For example, one-bead membrane models have been used to understand experimental results of temperature-mediated (large) liposome transport in hydrogels~\cite{yu_temperature-_2019}, to study vesicles~\cite{Noguchi2013,campos_dynamics_2020,azadbakht_wrapping_2023} and membrane-cytoskeleton interactions~\cite{jiang_modelling_2022}, to explore endocytosis and budding phenomena~\cite{azadbakht_wrapping_2023,Huang2013,weiner2024}, and even whole cell membrane reshaping~\cite{singh_bactericidal_2022,becton_flow_2019,harker-kirschneck_physical_2022,Fu2017,Appshaw2022}.
Among the different available models, the implementation of the YLZ one-bead model~\cite{Yuan2010} in the free, open-source software package LAMMPS~\cite{Fu2017, Thompson2022} makes it a particularly attractive choice for MD simulations of highly coarse-grained fluid lipid membranes. This model is applied in detail in section~\ref{subsec:ylz} to study membrane tube relaxation.

Membranes can be simulated with 1-bead models on length scales of $\sim 100 \mu \mathrm{m}^{2}$ and over timescales of up to minutes, making them suitable for modelling large biological membrane phenomena. However, these models have a few drawbacks. Although the ease with which one-bead-type membranes undergo scission and produce pores contributes to their success over mesh-based models, their rupture is also not always physical~\cite{wang_two-component_2024}. In fact, \textit{in vivo} or {in vitro} membranes do not typically cut as readily as these models would suggest~\cite{azadbakht_nonadditivity_2024}.
Additionally, implementing volume control within enclosed membrane surfaces in these models is not trivial.
Attempts to implement volume control using one-bead models include introducing coarse-grained solvent particles inside or outside membranes~\cite{campos_dynamics_2020,Fu2017} or using a triangulated volume~\cite{Yuan2010a}.
Nonetheless, the sensitivity and extent of volume manipulation in these methods is overall limited. Furthermore, as with mesh-based and continuum models, the bilayer thickness is not represented: the volume exclusion of the beads representing the membrane endows it with an artificial thickness, which makes these models unsuitable for looking at phenomena at scales $< 10 \mathrm{nm}$. Lastly, there is some discussion whether the kinetics of in-plane diffusivity versus out-of-plane mobility has the correct ratios in these models. This means, for example, that caution should be exercised where binding kinetics are being modelled alongside membrane deformations.
An attempt to rectify these inaccuracies by including partially correct hydrodynamics without introducing implicit solvent has been previously made~\cite{sadeghiLargescaleSimulationBiomembranes2020}.


%

\begin{figure}[t] 

    \begin{tcolorbox}[
        width=0.47\textwidth,          
        height=11cm,                
        colframe=white,             
        colback=cream,             
        arc=5pt,                  
        boxrule=1mm,               
        left=10pt, right=10pt,     
        top=10pt, bottom=10pt,
    ]

\textbf{BOX III: Conceptual guide}

\begin{enumerate}[leftmargin=1mm]
    \item []Below we present a series of questions that the reader might want to ask themselves to systematically develop a coarse-grained computational model that involves mesoscale membrane deformations.
\end{enumerate}

\renewcommand\labelitemi{\tiny$\bullet$}

\begin{enumerate}[leftmargin=2mm]
\item[1] \textbf{Relevant membrane properties}
\begin{enumerate}[leftmargin=2mm]
\item[-] What is the \textit{scale} of the deformation?
\item[-] Does the membrane \textit{thickness} play a role in the process?
\item[-] Is the membrane \textit{composition} homogeneous?
\item[-] Does the process involve a membrane \textit{topology} change?
\item[-] Is membrane \textit{dynamics} important?
\item[-] Should the membrane area or the enclosed volume be constrained?
\item[-] Are membrane hydrodynamics relevant?
\end{enumerate}
\item[2] \textbf{Computational implementation}
\begin{enumerate}[leftmargin=2mm]
\item[-] How are the external players, e.g. proteins represented?
\item[-] What are most appropriate boundary conditions for the system?
\end{enumerate}
\end{enumerate}

\end{tcolorbox}
\end{figure}

\section{Hands-on tutorial: Simulating mesoscopic membrane deformations with three model examples}\label{sec:tests}

In the first part of this section we provide the reader with some general advice for the development of mesoscale membrane simulations. We build on this in the hands-on tutorials of sections~\ref{subsec:trilmp}-\ref{subsec:cooke}, where we measure the force required to extrude a membrane tube using a tether with a mesh model, determine the equilibrium radius of a membrane tube as a function of membrane tension using a one-bead-per-lipid-patch model, and extract the fluctuation spectrum of a fluid membrane with a three-bead-per-lipid model.

\subsection{General advice for the development of mesoscale membrane simulations.}

By definition, a model is a simplified representation of reality. Deciding how to develop a model, or which model to use depends on (1) the system of interest, (2) the hypothesis to test, (3) the available computational resources and, potentially, (4) coding experience.
Points (3) and (4) are of a more mundane character and we hope to address them in sections~\ref{subsec:trilmp}-\ref{subsec:timing}. On the other hand, points (1) and (2) require more intellectual input from the modeller.
In this section we provide a conceptual guide to show the ways in which (1) and (2) can be addressed in a coarse-grained, mesoscale membrane model simulation.
A brief summary can be found in Box III, which suggests a series of questions to guide the reader in designing their computational coarse-grained model to study a biophysical problem featuring a fluid lipid membrane.

Given a specific problem, one must first determine which membrane properties must be accurately represented in the study the system of interest.
These properties depend both the scale of the whole phenomenon and on the scale of the membrane deformations.
Trivially, it can be too computationally costly to simulate large membranes using a fine-grained model (e.g. many beads per lipid); mesoscale properties might not have been benchmarked for such models either. Conversely, the membrane will not exhibit the correct behaviour if one tries to interpret fine details of membrane behaviour using a very coarse-grained model (e.g. triangulated mesh or one-bead-per-lipid-patch models). If the membrane thickness is thought to contribute in an important way to the studied process, mesh-based and one-bead-per-lipid-patch models are likely not suitable, as membrane thickness does not have physical interpretation in these models and these approaches do not capture any properties on the scale of a single lipid.

Another important consideration is in the representation of membrane dynamics. Is it sufficient to obtain a static (e.g. a free energy minimum) picture of the membrane shape, or does the phenomenon of interest has important dynamical features?
Membrane dynamics often comes into play through bending fluctuations and in-plane of lipid diffusivity, but it may also be important for some large membrane shape changes.
In these cases the continuum approaches should be disregarded in favour of dynamic mesh-based or particle-based models.
It should be noted that although mesh-based models can represent the in-plane fluidity of the membrane, the dynamics of the movement of mass is not captured. To properly capture lipid flow, one must turn to particle-based models (of appropriate coarse-graining level).
In processes where membrane fusion or fission plays a role, one may need to consider whether and at what level of detail topological changes are captured. As discussed above, mesh based models do not easily undergo topological changes whereas particle based models readily do, although some of them may not capture correctly the exact point of breaking or the two leaflets of bilayer membranes.

Other considerations include the overall state of the membrane, like constraints on the enclosed volume or membrane area. These are straightforward to include in mesh-based models and more challenging to capture in particle-based models.
Depending on the problem at hand, one needs to decide whether implicit or explicit solvent is more appropriate.
If small-scale dynamics and membrane fluctuations are of high importance, explicit solvent models could be more accurate, while implicit solvent models result in simulations which are orders of magnitude faster.
Likewise, most membrane remodelling processes involve non-lipid external players. For example, to study the interactions between cell membranes and cytoskeleton filaments one may represent the filament-membrane interactions as a point forces on the membrane, or, if higher resolution is desired, filament geometry may be included, e.g. by representing it using bonded particles interacting with the membrane.
Other ways to induce membrane deformations may be to include specific boundary conditions or to use global force fields in the simulation.

Despite all these considerations, it is important to note that often several distinct approaches can be applied the same biophysical system. This is highlighted in Fig.~\ref{tab:Table1}, and discussed in the following sections, where we describe three classical simulation tests using three different membrane models.

\subsection{Force required to extrude a membrane tube with a triangulated-mesh model}\label{subsec:trilmp}

\begin{figure*}
    \centering
    \includegraphics[width=18.1cm]{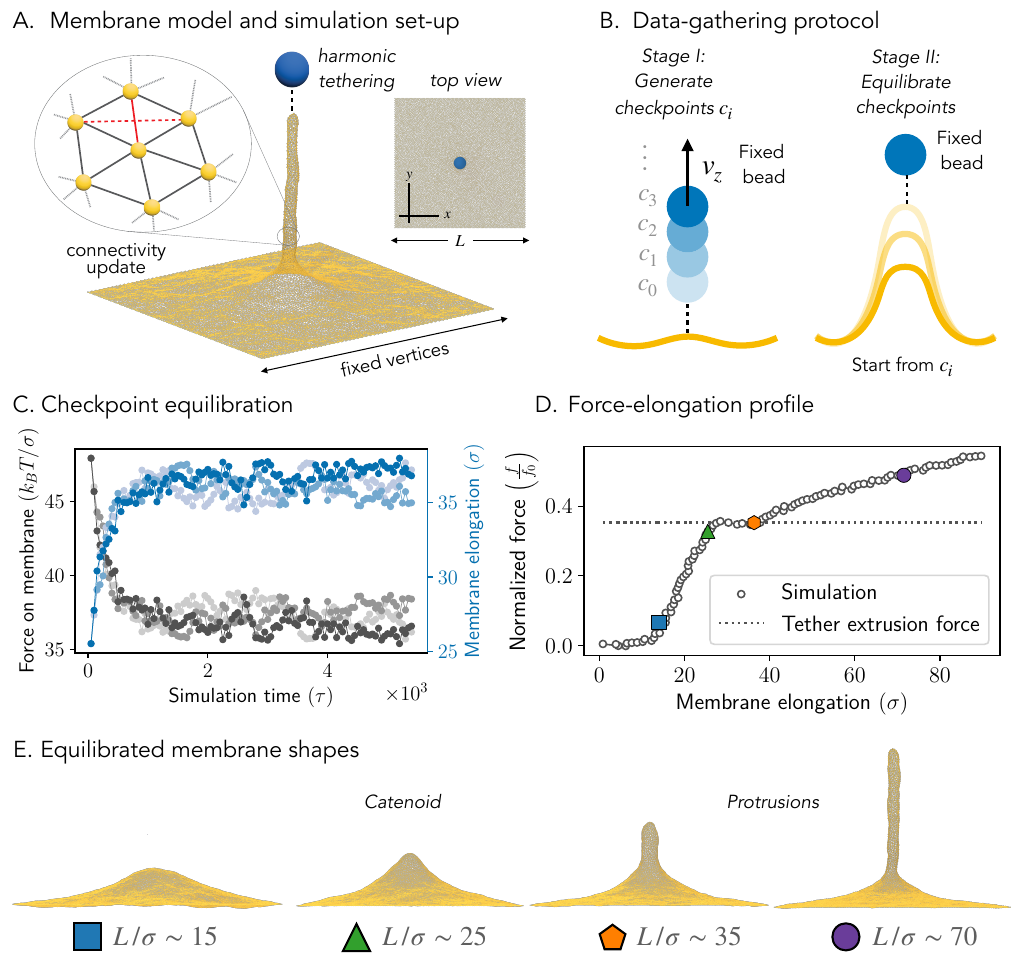}
     \caption{Tether pulling simulation using a mesh model. (A) In a dynamically triangulated mesh, membrane fluidity is ensured by regularly swapping bonds in the mesh. To extrude a membrane tube, we set-up a flat membrane patch, fix the vertices at the edges of the patch and tether the central membrane vertex to a bead (blue particle) via a harmonic spring. (B) To measure the force required to extrude a tether, we develop a two-stage protocol: first, we move the blue particle with constant velocity $v_z$ to generate simulation checkpoints; second, we reinitialize the system at each checkpoint to let the membrane relax for a fixed position of the blue bead. (C) Example of relaxation curves for the force pulling on the membrane (shades of black) and membrane elongation (shades of blue) obtained during the second protocol step for a membrane with $\kappa = 20 k_\mathrm{B} T$ and $\gamma = 7 k_\mathrm{B}T /\sigma^2$ using TriLMP~\cite{TriLMP}. The blue bead is fixed at $z \approx 48\sigma$ from the initial membrane plane. As the simulation progresses, the pulling force on the membrane equilibrates to a non-zero value. (D) Force-elongation profile. The curve is obtained by averaging over the final equilibrated force in three replica simulations. The standard error on the mean is smaller than the data-points. The force $f$ measured in simulations is normalized by $f = 2\pi \sqrt{2\kappa \gamma}$. (E) Representative simulation snapshots for different membrane elongations shown as colored points in panel D.
    \label{fig:elasticnetwork}}
\end{figure*}

As tube extrusion using a tether is relatively straightforward to perform both experimentally and in computer simulations, it has been extensively used to mechanically characterize lipid membranes~\cite{Bassereau2014,koster_force_2005, paraschiv_influence_2021, weichsel_more_2016}.
In this section we describe a simulation protocol to study the formation of a membrane tether using a dynamically triangulated mesh.
Specifically, we perform our simulations using TriLMP~\cite{TriLMP}, an in-house software that couples the TriMEM~\cite{siggel2022} mesh-based software with the LAMMPS~\cite{Thompson2022} MD simulation package. TriLMP is available as an online git repository~\cite{TriLMP}. The protocol we describe in this section is general and can be easily used with any other triangulated mesh model~\cite{matoz-fernandez_pymembrane_2023, zhu2022,dadunashvili_flippy_2023}.

\subsubsection{Simulation set-up}
Membrane tubes extruded by pulling a tether have been showed to follow predictions of the Helfrich analytic theory, namely, a radii of $R_\mathrm{eq}=\sqrt{\kappa/(2\gamma)}$ and a required force for tether extrusion $f = 2\pi \sqrt{2 \kappa \gamma }$ \cite{Derenyi2002}.
Using typical numbers for bending rigidity $\kappa$ and membrane tension $\gamma$, we expect tube radii below optical resolution, ranging from tens to hundreds of nanometres.
Therefore, the membrane area that moves into the tube is small in comparison to the total area of the cells and vesicles they are experimentally extruded from.
For this reason, it should be sufficient to restrict our simulation to a membrane patch rather than simulating a full vesicle, as this is the scale relevant for the deformation~\cite{weichsel_more_2016}.

To conduct the simulation, we initialize the system as a square, flat triangulated mesh (Fig.~\ref{fig:elasticnetwork}A), which was generated by arranging a collection of points as a two-dimensional triangular lattice and connecting the nearest-neighbour vertices.
The number of vertices $N_V$ must be sufficiently large to avoid finite size effects, and is governed by the trade-off between simulation resolution and the computational time. In this tutorial  $N_V = 11500$ vertices.
The vertices that belong to the edge of the mesh will remain fixed during the simulation (Fig.~\ref{fig:elasticnetwork}A).
Fixing the edges of the membrane patch implies that the projected area $A_\mathrm{P}$, i.e. the area of the membrane projected onto the surface without fluctuations, remains constant throughout the simulation.
The value of $A_\mathrm{P}$ depends on the initial average mesh bond length $\bar{l}$.
Triangulated mesh models constrain the length of the edges in the mesh between a maximum $l_{\max}$ and minimum $l_{\min}$ elongation to prevent the mesh from self-intersecting.
The minimum elongation $l_{\min}$ can be interpreted as the diameter $\sigma$ of a particle placed on a mesh vertex, while $l_{max}$ satisfies $l_{\max}/\sigma = \sqrt{3}$.\cite{gompper_triangulated-surface_2004}
The choice of $\sigma$ sets the length scale of the system.
In addition, $\bar{l}$ is chosen such that $A\gg A_\mathrm{P}$, where $A$ is the total membrane area including fluctuations. This means that the membrane patch will have sufficient \textit{excess area} to extrude a tube.
When we update the positions of the vertices in the mesh throughout the simulation, the area of each triangle in the mesh changes.
In the example presented in Fig.~\ref{fig:elasticnetwork}, we set $\bar{l}/\sigma = 1.05$, which leads to a side of the simulation box $L/\sigma \approx 100$ and an initial projected area of $A_\mathrm{P}/\sigma^2 \approx 10775$.

Once the initial conditions of the simulation are set, we must define the membrane energy function.
This will dictate how the vertices in the mesh fluctuate and respond to deformations.
The energy function used to describe the membrane in TriLMP,
is based on a discretized version of the Helfrich bending energy~\cite{helfrich1973, siggel2022}, typically together with contributions that penalize area increase and constrain how much can the enclosed volume change.
We are simulating a flat membrane patch, so no volume constraint applies, and we only need to set the bending rigidity, which we choose as $\kappa = 20~k_\mathrm{B} T$~\cite{Bassereau2014} and the membrane tension, which we set to $\gamma \sigma^2 = 7$.
The value of the membrane tension parameter $\gamma$ is chosen such that the pinned membrane patch, 
remains roughly flat as it fluctuates in the prescribed set-up.
Smaller values of $\gamma$ lead to the spontaneous emergence of finite size effects such as membrane buckling and bending due to the fixed boundary conditions.

\subsubsection{Integrating the dynamics of the membrane}

TriLMP simulates the dynamics of a fluid membrane using an HMC approach, that is, by combining MC and MD stages throughout the integration. Specifically, a typical simulation run consists of a sequence of two stages: bond-swaps, which are determined by an MC algorithm (Metropolis criterion), and vertex moves, which are performed using MD. We can choose whether we want the order of MD and MC stages to occur at random throughout the simulation (our choice for the tutorial), or whether we prefer to alternate between the two stages. During the bond-swap stage, the vertices in the network remain immobile, and we try to sequentially update a maximum prescribed fraction $f_\mathrm{s}$ of the edges in the network (Fig.~\ref{fig:elasticnetwork}). All edges in the network can be swapped throughout the simulation, and the choice of $f_\mathrm{s}$, as well as the length of an MD stage impact the fluidity of the membrane. Large values of $f_\mathrm{s}$ imply a large number of bond swap attempts during an MC stage at the expense of increasing computational time. We found a good trade-off to be the value of $f_{s} = 20\%$. During the MD stage of the simulation, the positions of all vertices in the network are updated simultaneously.

The motion of the vertices in the membrane, with mass $m$, is evolved in time using the velocity-Verlet scheme.
The length of the MD stages is set to  $N_\mathrm{MD}$ time steps and it
determines how much the vertices of the mesh will move before we attempt to update the network connectivity.
Thus, just like $f_\mathrm{s}$, $N_\mathrm{MD}$ has an impact on membrane fluidity.
In our case, we set $N_\mathrm{MD} = 50$ steps, which, in combination with the choice of $f_\mathrm{s}$ discussed above, leads to a satisfactory dynamics.
Choosing a timestep of $\mathrm{d}t = 0.001$ to integrate the equations of motion implies that the length of the MD stages is $\tau_\mathrm{MD} = N_\mathrm{MD} \times \mathrm{d}t = 0.05\ \tau$, where $\tau$ is the simulation units of time.
Additionally, it is at this stage of the simulation where we decide whether the membrane will be in contact with an explicit or implicit solvent. In our case, we chose an implicit solvent approach, namely, a Langevin thermostat at temperature $T$ (with $k_\mathrm{B} T = 1$ as energy scale for the system), as it is computationally cheaper and sufficient for our purposes.
The thermostat, applied to the membrane using the LAMMPS implementation, is coupled to the system with a damping coefficient $\mathtt{damp} = 1 \tau$ (see the LAMMPS documentation for details).
Finally, we note that having set the scales for length, $\sigma$, mass, $m$, and energy, $k_\mathrm{B}T$ in the system, the time scale $\tau$ is given by $\tau = \sigma \sqrt{m/k_\mathrm{B} T}$.

\subsubsection{Protocol to measure the extrusion force}

Mimicking the optical tweezer experiments~\cite{koster_force_2005, Bassereau2014}, the desired membrane deformation can be achieved by tethering a bead to the membrane patch and pulling on it (Fig.~\ref{fig:elasticnetwork}A).
In our tutorial, we use a bead that is tethered to the central vertex of the membrane through a harmonic bond.
It is important to choose the elastic constant of the bond to be weak, so that one can simulate adiabatic tube extrusion~\cite{paraschiv_influence_2021}.
Here we use $\sigma_\mathrm{B} = 10\sigma$ for the diameter of the bead, $k = 1 k_\mathrm{B} T/\sigma^2 $ as the elastic constant of the bond and set the rest length of the bond to $\tilde{\sigma} = 1/2 (\sigma + \sigma_\mathrm{B}) = 6.05~\sigma$.

To study the extrusion force in a simulation, we develop a two-step protocol that allows us to equilibrate the system and obtain the force-elongation profile in an efficient manner (see Fig.~\ref{fig:elasticnetwork}B).
We first run a long simulation where the bead tethered to the membrane is displaced along the direction perpendicular to the membrane patch.
We do so with constant velocity of $v_z = 0.1 \sigma/\tau$, which is large enough to observe the formation of tubes fast, but sufficiently slow to allow the diffusion and flow of vertices as the membrane deforms.
As the bead moves, we record simulation checkpoints at a given frequency.
We next reinitialize new simulations from all recorded checkpoints.
In these new simulations, the bead position is fixed and we allow the membrane to relax. Fig.~\ref{fig:elasticnetwork}C shows an example of a set of relaxation curves when the pulling bead is fixed at some $z = z_0$ above the initial plane of the membrane.
The membrane equilibrates by elongating towards the pulling bead as the simulation progresses, which minimizes the energy stored in the harmonic bond that connects them.
For a given bead position, the force required to deform the membrane is calculated from the deviation of the harmonic bond from its rest length once
the system has equilibrated.

\subsubsection{Results: analyzing the force-elongation profile}

The force-elongation profile during tether extrusion of a fluid membrane can be divided into several regimes that have been experimentally measured~\cite{koster_force_2005, paraschiv_influence_2021}.
At small elongations, one finds an elastic regime where the force required to deform the membrane is proportional to the elongation; here the membrane adopts the shape of a catenoid~\cite{powers_fluid-membrane_2002}.
At larger deformations, the profile may exhibit an overshoot~\cite{koster_force_2005, Derenyi2002}, followed by a saturation regime where the force plateaus.
The plateau indicates the collapse of the catenoid into a tube with the radius $R_\mathrm{eq} = \sqrt{\kappa/2\gamma}$.
While theoretical calculations often assume the existence of a reservoir of lipid molecules, and thus, the possibility of increasing area without changing membrane tension~\cite{Derenyi2002, powers_fluid-membrane_2002}, in physical systems a new elastic regime can emerge indicating the depletion of excess area available for the tube.

The curve elongation-profile obtained in our simulations is shown in Fig.~\ref{fig:elasticnetwork}D.
This is computed by extracting the membrane equilibrium elongation value, which we take as the position of the membrane vertex that the pulling bead is bonded to, and the equilibrium force that the membrane exerts on the bead, for different simulation checkpoints.
As described above, the curve has first an elastic regime where the shape of the membrane is catenoidal, followed by a plateau where the tube forms (Fig.~\ref{fig:elasticnetwork}E).
We also capture a second elastic regime at large elongations due to the finite area in the simulated membrane patch.
This limitation can be circumvented by performing grand-canonical simulations, where the membrane is coupled to a lipid reservoir and thus can grow, extending the plateauing region in the force-elongation profile~\cite{weichsel_more_2016}.
We finally note that the results in Fig.~\ref{fig:elasticnetwork}D do not exhibit the plateauing at the expected theoretical value of $f$.
The discrepancy between theory and simulations can be rationalised by arguing that the assumptions that the theoretical calculation relies on are not fulfilled in the simulations.
In particular, aside from not being coupled to a lipid reservoir, the projected area $A_\mathrm{P}$ of our membrane patch is kept constant at no cost by pinning the edges of the membrane.



\subsection{Membrane tube equilibration using the YLZ potential}\label{subsec:ylz}
In this section, we simulate membrane tubes of varying bending rigidity at different tensions using the YLZ model~\cite{Yuan2010}, and compare the simulation results to the analytic theory of section~\ref{subsubsec_continuummodels}. We note that while membrane tube equilibration could be studied using the same procedure as described in section~\ref{subsec:trilmp} for TriLMP, here we detail an alternative approach that exploits periodic boundary conditions and a barostat to control in-plane membrane tension.

\subsubsection{The YLZ potential}
One of the most widely used 1-bead-per-lipid-patch models is the YLZ model; developed in 2010 and named after its authors\cite{Yuan2010}.
This set-up models a biological membrane as a single layer of particles interacting via a 2-body potential dependent on the relative distance and orientation of the particles (Fig.~\ref{fig:YLZ}A).
When this potential is implemented in MD simulations, beads self-assemble into vesicles, tubes or sheets, and exhibit membrane properties such as biologically relevant fluidity and bending rigidity, which can be tweaked via the bead-bead potential parameters.
In the simulations presented in this section, we used the values of YLZ parameters given in Table~\ref{tab:ylz_params}, unless stated otherwise.
How these parameters relate to the underlying potential can be seen in the original article introducing the potential~\cite{Yuan2010} or the LAMMPS documentation section \texttt{pair\_style~ylz}.

\begin{table}[]
    \centering
    \begin{tabular}{@{}cc@{}}
        \toprule
        \textbf{Parameter}                  & \textbf{Value}\\ \midrule
        $\varepsilon $   & $4.34$                               \\ 
        $\mu$ & $2.5, 3, 3.5$                 \\ 
        $\zeta$                      & $4$ \\ 
        $r_\mathrm{min}$                      & $\sqrt[6]{2}$     \\
        $r_\mathrm{c}$   & $2.6$               \\ 
        $\theta_0$                        & $0$                                  \\ \bottomrule
    \end{tabular}
    \caption[Table caption text]{Parameters used for the YLZ potential. All values are in simulation units.}
    \label{tab:ylz_params}
\end{table}

\subsubsection{Simulation set-up and protocol to measure the equilibrium radius}
To set up the initial state, we put YLZ particles on a regular trigonal lattice on the surface of a cylinder with the edge length roughly corresponding to the minimum of the radial potential ($\approx 1 \sigma$, where $\sigma$ sets the lengthscale of the system).
The orientation vectors are all taken to point outside of the tube, normal to the surface of the cylinder.
The simulation box is chosen such that the tube ends join onto themselves via the periodic boundary conditions.
The other two dimensions of the box are irrelevant, as long as we the tube does not interact with images of itself at any point in the simulation (Fig.~\ref{fig:YLZ}A).

To measure the equilibrium tube radius, we use the following protocol. The tube is first equilibrated to a constant temperature at a constant volume using the standard velocity Verlet algorithm in combination with the Langevin thermostat as implemented in LAMMPS.
The tube then relaxes and adopts an average radius corresponding to a specific tension as given by the theory $R_\mathrm{eq}=\sqrt{\kappa/(2\gamma)}$.
Measurement of the stress tensor can provide the value of this tension as shown in reference~\citenum{Harmandaris2006}.
In the next phase of the simulation, the membrane tube is equilibrated under the action of a Nose-Hoover barostat as implemented in LAMMPS. The LAMMPS code has been modified to redefine the pressure estimator, as outlined in the next section.
The radius of the equilibrated tube is then measured by fitting a circle to its cross section.

\subsubsection{Regulating membrane tension in a simulation}
The main conceptually difficult part of the simulation is the use of a barostat to control the tension of the membrane.
As can be derived  from the definition of a surface tension, the membrane tension of a tube can be related to its equilibrium radius and the force in the direction of the tube axis exerted by the membrane as
\begin{equation}
    \gamma = \frac{f_z}{4\pi R_{\mathrm{eq}}}.
\end{equation}
The same formula can also be obtained from the Helfrich Hamiltonian, as done in \cite{Harmandaris2006}. From the definition of the pressure and stress tensors (in the convention used by LAMMPS), the same force can be obtained as
\begin{equation}
    f_z = -P_{zz} L_x L_y = \frac{\sigma_{zz}}{L_z}.
\end{equation}
where $P_{ij}$ is the pressure tensor, $\sigma_{ij}$ is the stress tensor and $i,j$ refer to the three dimensions $x,y,z$.
To control membrane tension with a barostat, we defined a quantity $P'$ that can be used by the barostat and that is easily relatable to the membrane tension. This quantity is defined as
\begin{equation}
    P'_{zz} = \frac{P_{zz}V}{N} = -\frac{\sigma_{zz}}{N},
\end{equation}
where $N$ is the number of the membrane particles. This is basically the negative of the average stress per YLZ particle.
We can relate $P'$ to $\gamma$ as
\begin{equation}
    \gamma = - \frac{P'_{zz} N}{4 \pi R_{\mathrm{eq}} L_z} = - \frac{P'_{zz}}{2a},
    \label{eq:sigma_from_YLZ_simulation}
\end{equation}
where we used the area of the membrane cylinder as $A = N a = 2\pi R L_z$, with $a$ being the membrane area per YLZ particle.
This allows us to calculate the value of the membrane tension for our simulations.
In addition, if we assume that $a$ is approximately independent on the tension, then $\gamma \propto -P'_{zz}$.
Empirically, the membrane area per YLZ particle has not varied over our simulations more than $\pm 5 \%$.
However, since the area per YLZ particle can be also obtained from the simulation, one does not have to rely on it being constant for the calculation of tension.

\subsubsection{Comparison with theory and computation of the bending rigidity $\kappa$}
Informed by the analytical formula for $R_\mathrm{eq}(\gamma)$, we plot $\ln R_\mathrm{eq}$ as a function of $\ln \gamma$ and observe that these two quantities adhere extremely well to a power law, which can be represented as
\begin{equation}
    R_\mathrm{eq} = \left(\frac{\kappa}{2\gamma}\right)^{x}
\end{equation}
The equation fitted to the data in Fig.~\ref{fig:YLZ}D is
\begin{equation}
    \ln R_\mathrm{eq} = x \ln \frac{\kappa}{2} - x \ln \gamma
    \label{eq:ylz_fit}
\end{equation}
The simulations are consistent in the value of $x$ (see Fig~\ref{fig:YLZ}D), but slightly deviate from the theoretical value of $1/2$ (see Fig.~\ref{fig:YLZ}E). We varied $\mu$, the parameter controlling the bending rigidity, and calculated the bending rigidity $\kappa$ for each value using equation~\ref{eq:ylz_fit} (see Fig.~\ref{fig:YLZ}D), which we compared to the $\kappa$ values obtained in reference~\citenum{Yuan2010} through analysis of fluctuation spectra.

\subsubsection{Results: The effects of the tension $\gamma$ and the bending rigidity $\kappa$}
\begin{figure*}
    \centering
    \includegraphics[width=17.1cm]{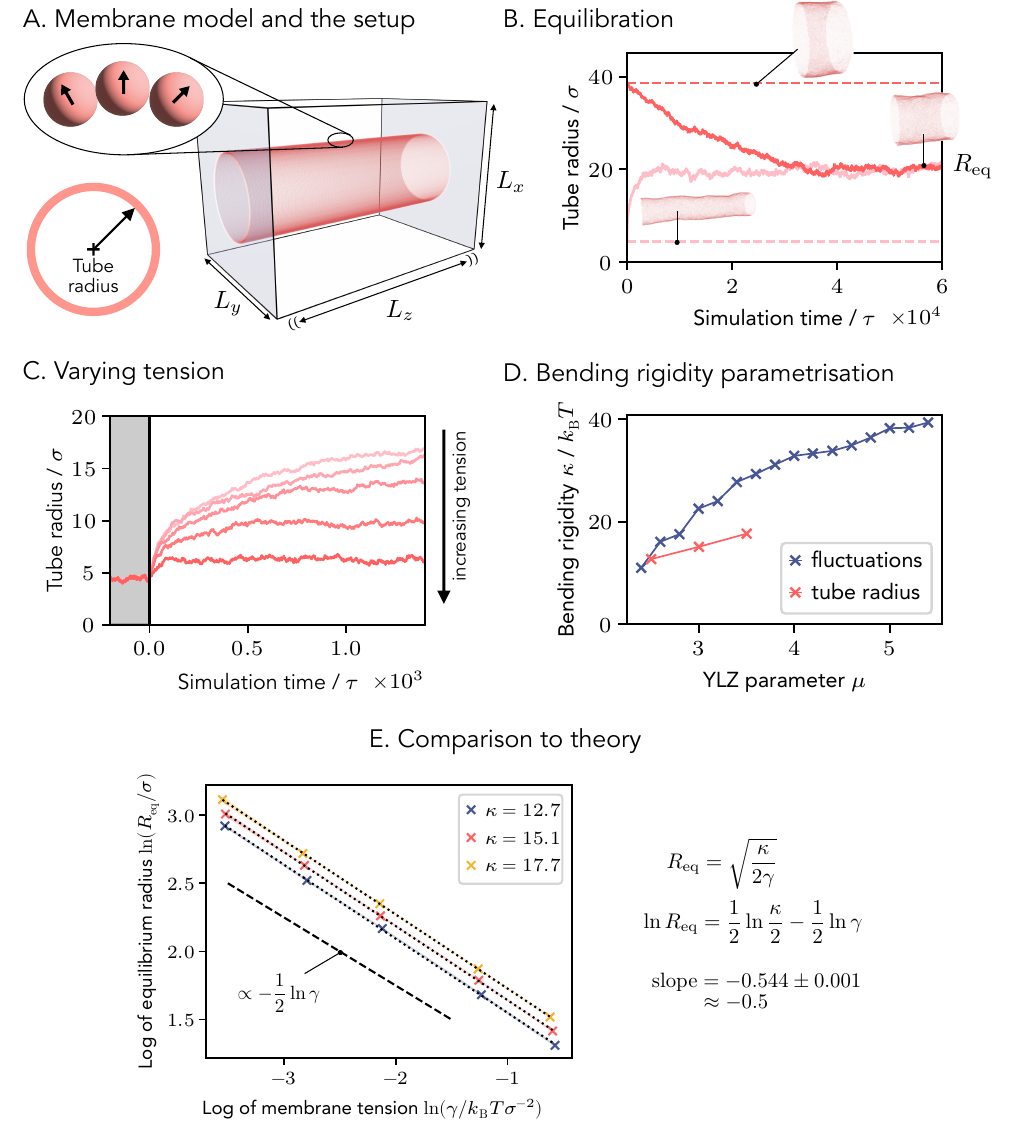}
    \caption{\textbf{Membrane tube equilibration using the YLZ potential} A) The YLZ model models the membrane as a single layer of interacting beads. A tube of an arbitrary radius is constructed out of a regular trigonal lattice in a simulation box with periodic boundary conditions. B) The size of the box along the cylinder axis is allowed to vary using a modified NPH Nose-Hoover barostat. Regardless of the initial state, under the action of the barostat, the membrane always equilibrates to the same radius if we set the tension and membrane parameters to be the same. C) As predicted by the theory, the same membrane tube equilibrates to a different radius under different tensions with radius decreasing with increasing tension. (The shaded part shows a fixed-box equilibration.) D) The parameter $\mu$ of the YLZ model changes the bending rigidity of the membrane, $\kappa$. The value of $\kappa$ was estimated for three different values of $\mu$ using the tube radii at different tensions. This is compared to the estimates from the membrane fluctuation spectrum taken from \citet{Yuan2010}. E) The Helfrich theory predicts that the radius of a thermodynamically stable tube is proportional to $1/\sqrt{\gamma}$. The simulations adhere to a power law nearly perfectly, with the exponent being close to that given by the theory. The equilibrium value of the radius for each set of parameters was obtained by averaging over time and three different random seeds.
    The three seeds were also used for an estimation of the error, which, however, is negligible (see the barely visible shaded area).All quantities in the figure are expressed in simulation units.}
    \label{fig:YLZ}
\end{figure*}

The simulations were stable except under very high tensions and occasionally shedding a small fraction of the particles (which can be solved by adjusting barostating parameters). The results agree with the theory extremely well. For the same set of membrane parameters and the same membrane tension ($\approx$ same $P'_{zz}$), tubes equilibrate to the same equilibrium radius regardless of their initial radius and length (thus also the number of particles) (see Fig.~\ref{fig:YLZ}B). The equilibrium radius decreases with increasing membrane tension (Fig.~\ref{fig:YLZ}C) and the simulations closely follow $\approx-1/2$ dependence as predicted by the theory (Fig.~\ref{fig:YLZ}E). The membrane bending rigidity $\kappa$ is set by the parameter $\mu$ and the equilibrium radius increases with increasing rigidity as expected. The same dependence was measured previously by reference \citenum{Yuan2010} using the fluctuation spectrum. The results are noticeably different, which higlights that the YLZ potential is not a mere numerical solution of the Helfrich Hamiltonian (eq.~\ref{eq:Helfrich}), but rather a coarse-grained potential mimicking the membrane properties in a less straightforward way.

\begin{figure*}
  \centering
    \includegraphics[width=17.1cm]{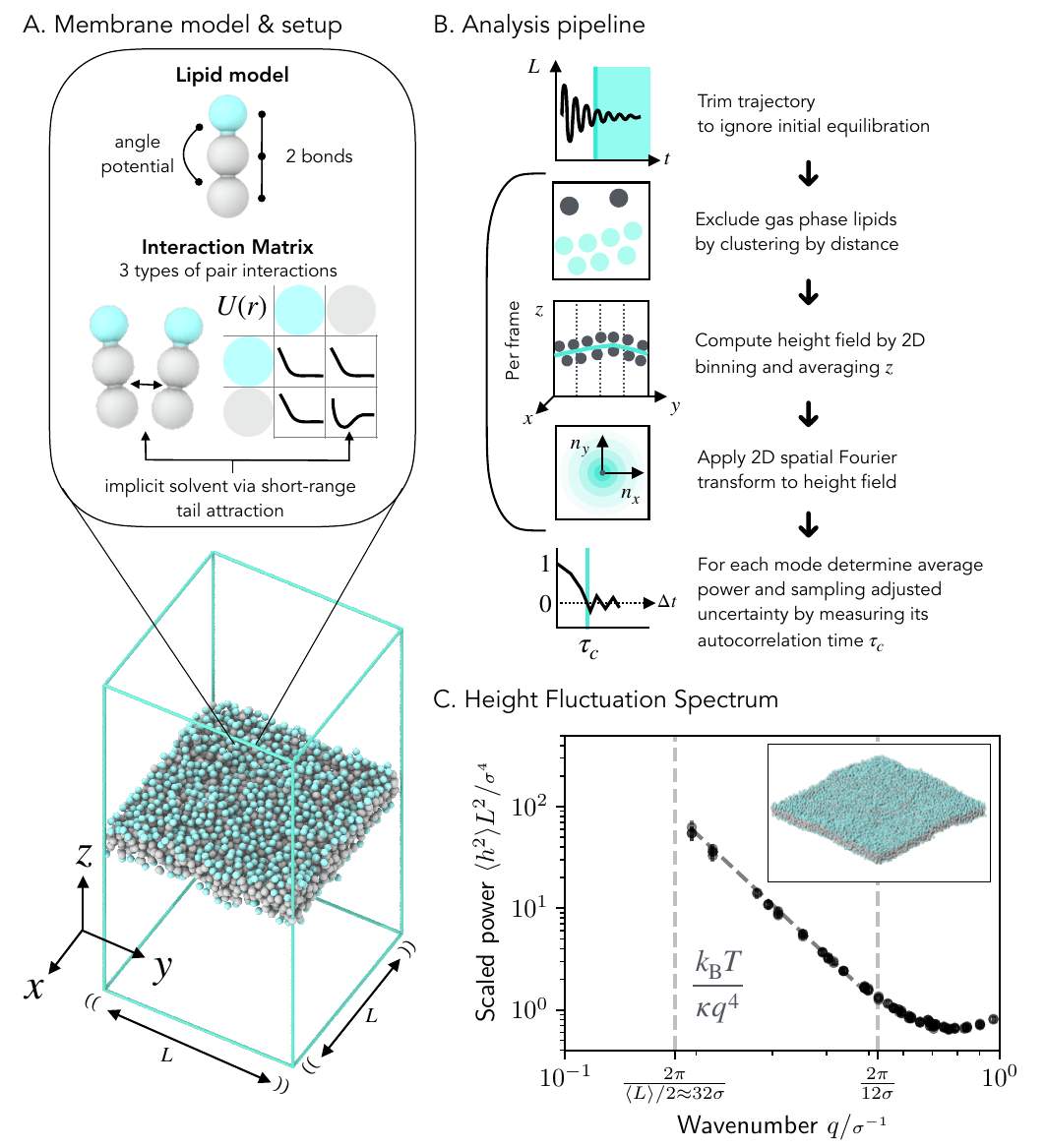}
  \caption{
    (A) (top) Diagram of Cooke bilayer lipid and interaction matrix showing pair potential sketches.
    (bottom) Simulation setup for a small membrane patch (small size chosen for clarity), with double arrows showing the barostat-coupled $x,y$ box dimensions $L_x=L_y=L$.
    (B) Analysis pipeline: skip over initial equilibration determined by the evolution of the box length $L$ in time, then for the rest of the trajectory consider $L$ to be $\langle L \rangle$; process equally spaced frames to obtain for each the instantaneous height fluctuation spectrum; finally for each mode of the spectrum compute its average power $\langle h^2 \rangle$ and corresponding uncertainty as the adjusted standard deviation $\sqrt{\mathrm{Var}(h^2)\frac{2\tau_c/\tau+1}{N}}$.
    (C) Ensemble average the height fluctuation power spectrum, in log-log space, with snapshot of simulation.
    In the interval where behaviour is linear (wavelengths between $32\sigma$ and $12\sigma$, marked by vertical lines), we do a linear fit (dashed grey line).
  }
  \label{fig:test_cook_spectrum}
\end{figure*}

\subsection{Analysing membrane fluctuation spectrum using the Cooke model}
\label{subsec:cooke}

A lipid membrane is thin and soft enough for the thermal energy to manifest as shape fluctuations. In this section, we describe a simulation setup to measure height fluctuation spectra using the coarse-grained Cooke \cite{Cooke2005} model for lipid membranes and LAMMPS.

For a flat, tension-less square membrane patch of side of length $L$ and on the $xy$ plane, shape fluctuations result in height fluctuations in the $z$ direction.
For small deformations, one can use the small deformation approximation known as the Monge Gauge, where the surface of the membrane is defined by a varying offset $h(x,y)$ with respect to a flat plane and model the ensemble spectrum of $h(x,y)$.
As introduced in section~\ref{subsubsec_continuummodels}, the mean square amplitude of the mode with number $n$ follows
\begin{equation}
 \langle h_q^2 \rangle  = \frac{\kBT}{\kappa L^2 q^4},
 \label{eq:cooke_spectrum}
\end{equation}
where $q=2\pi n /L$ is the wave number.
Eq.~(\ref{eq:cooke_spectrum}) shows that shape fluctuations provide a connection between the scale of the thermal deformations and the bending modulus.

\subsubsection{The three-beads-per-lipid model}
One of the most recent and widely used several-beads-per-lipid models is the Cooke membrane model~\cite{Cooke2005}, where the solvent is modelled implicitly. We refer the reader to its seminal paper\cite{Cooke2005} for specifics, and give only a brief overview here.

In the model, each lipid consists of a three-bead-long chain, a head bead $b_1$ followed by two tail beads, $b_2,b_3$ (see Fig.~\ref{fig:test_cook_spectrum}A).
The bonds are Finite Extensible Nonlinear Elastic (FENE) bonds, i.e. they keep consecutive beads at near fixed distance, and the lipids are kept straight by a harmonic angle potential on the angle formed by
$\angle{b_1 b_2 b_3}$\cite{Cooke2005}.
We use $\sigma$ as distance unit, $\tau$ as the time unit, and $\epsilon$ as energy unit.
All beads repel each other via a purely repulsive Lennard-Jones potential that becomes zero when particles are $1\sigma$ apart.
To represent the solvent implicitly, the hydrophobic interaction between tails is modelled as an added attractive potential with a tunable range $w$ between tail beads of different lipids.
The tail to tail attractive interaction compresses the tail region, and would lead to spontaneous curvature without further modifications, which for a minimal system introduces extra terms in the Hamiltonian and is thus undesirable.
This is why, for interactions with head beads, the purely repulsive potential is adjusted to become zero at $0.95\sigma$, and thus why we render head beads with a slightly smaller radius~\cite{huDeterminingGaussianCurvature2012}.
As a consequence, lipids self-assemble into bilayer membranes.
Importantly, the rate of individual lipids switching between the bilayer membrane leaflets (flip-flopping) is much higher than reported in experimental values~\cite{Cooke2005}, which on the other hand facilitates computing equilibrium properties by making membrane shape relaxation faster.
The Cooke models is typically limited to length scales of $\sim \mu \mathrm{m}^{2}$ and timescales of seconds.

\subsubsection{Simulation set-up}
The seminal paper for the Cooke model \cite{Cooke2005} tuned two parameters, the tail interaction range $w$ which we set to $1.5\sigma$, and the temperature which we set up so that $\kBT=1\epsilon$; this guarantees that our system will be in the liquid phase.
We also used the same timestep of $\mathrm{d} t=0.01\tau$.
To simulate this system, we pre-assemble a flat membrane in a periodic box with dimensions $(L,L,L_z)$, where we pick $L=60\sigma$.
We chose to place each lipid molecule in a hexagonal grid with two layers, one per leaflet, oriented so that their head beads point away from the membrane.
We then setup a Noose-Hover NPH barostat, with the relaxation constant of $10\tau$, whose function is to scale the simulation box length $L=L_x=L_y$ to enforce zero lateral pressure $P_x,P_y=0$.
This is so that the membrane is neither stretched nor compressed, or equivalently, so that membrane tension is kept zero, although we must stress that this is only partially accurate, since the membrane will fluctuate, and thus the membrane will not exactly be aligned with the simulation box horizontal $xy$ plane.
To keep the system thermalized at constant temperature, we also setup a Langevin thermostat with the relaxation constant $\mathtt{damp} = 1\tau$.
We picked the duration of our simulation to match that of Cooke's : $10\cdot10^3\tau$ reserved for initial equilibration, followed by $50\cdot 10^3\tau$ for measurements, for the total of $60\cdot10^3\tau$.
Additionally, to be able to distinguish what might seem like an equilibrated observable from one that is kinetically trapped or varying too slowly to be observed, we ran four replicas of the simulation with different seeds of the random number generator used by the thermostat to draw velocities.
We setup simulation output so that every $100\tau$ we record the average potential energy, the global temperature and pressure for each box axis, the simulation box dimensions, and the particle coordinates.

\subsubsection{Analysis pipeline}
We generally follow the analysis provided in \cite{Cooke2005}, informed by a latter work which focuses on fluctuation spectrum analysis \cite{erguderIdentifyingSystematicErrors2021}.
Fig.~\ref{fig:test_cook_spectrum}C provides an overview of our analysis.
First, we determine if the equilibration stage is complete by plotting together the 4 replicas.
We verify qualitatively that the simulation has equilibrated within the initial $10^4\tau$ of the simulation by monitoring the box length $L$ rather than temperature or energy of the system, both of which equilibrate much faster.
By visually rendering the simulation, we verify the integrity of the membrane for the rest of the trajectory since lipids can occasionally evaporate and be adsorbed by the membrane.
For each trajectory frame, we then apply three transformations.
First, we cluster the simulation particles, grouping together particles that were within $1.5\sigma$ of each other, and then take the largest cluster as the membrane and exclude the rest, thus removing the gas phase.
Secondly, we compute the membrane height field $h_{ij}$ by computing the 2D histogram of the particles $z$ coordinates with bins covering the $xy$ section of the simulation box, and then taking the average height $h_{ij}=\langle z \rangle_{ij}$ in each bin.
Here we pick the number of bins per axis to be both a power of two (for a faster Fourier transform), and such that each bin square at least covers $3^2\sigma^2$, which for $L=60\sigma$ yields 16 bins per axis.
Finally, we take the 2D discrete real Fourier transform of $h_{ij}$, obtaining $H_{ij}$.
We note that while $L$ is not fixed during our simulation, the theory of height fluctuation spectrums expects a fixed $L$. Nonetheless, because the changes in $L$ are small, we simply use the average $\langle L \rangle$ when fitting the spectrum.
For each mode, identified by its 2D wave number $\vec{n}=(i,j)$, we can then compute the average and standard deviation of the squared amplitude, $\langle\left|h^2\right|\rangle$ and $\operatorname{Var}\left(\left|h^2\right|\right)$.
Because we are interested in estimating the error of the mean value, we then analyse the time series of the complex amplitude to determine its correlation time.
To do this, we compute the statistical inefficiency $g_\mathrm{ineff}$ (see \cite{choderaSimpleMethodAutomated2016}) for its norm squared $|h|^2$ and phase $\angle h$, taking the largest of the two as the effective $g_\mathrm{ineff}$.
If $N$ is the number of points in the time series, the effective sample side, i.e. the number of uncorrelated observations, is $N/g_\mathrm{ineff}$. The error of the mean is then $\operatorname{Var}\left(\left|h^2\right|\right)/g_\mathrm{ineff}$.

\subsubsection{Results: analysing membrane fluctuations}

In Fig.~\ref{fig:test_cook_spectrum}C, we show the membrane height fluctuation spectrum.
For the membrane parameters and the simulation duration used, modes are well-sampled only if their wavenumber $n\geq2$, or equivalently, if their wavelength is smaller than half the length of the simulation box.
We also observe that the initially linear trend becomes nonlinear, with the amplitude increasing above the expected value.
While it is possible to try to pick the maximum $q$ threshold based on the fitness of the fit, here we took a simpler approach of setting it so the equivalent wavelength is $12\sigma$, or roughly twice the membrane thickness.
By fitting the spectrum in the initial linear region to Eq.~(\ref{eq:cooke_spectrum}), we can obtain the bending modulus $\kappa=11.4\pm0.1\kBT$.
We note that the key bottleneck in this measurement is the sampling of the mode with smallest wave number of interest because the relaxation time of a mode scales as $q^{-4}$.~\cite{Cooke2005,farago2008}.


\begin{figure*}[ht!]
    \centering
    \includegraphics[width=17.1cm]{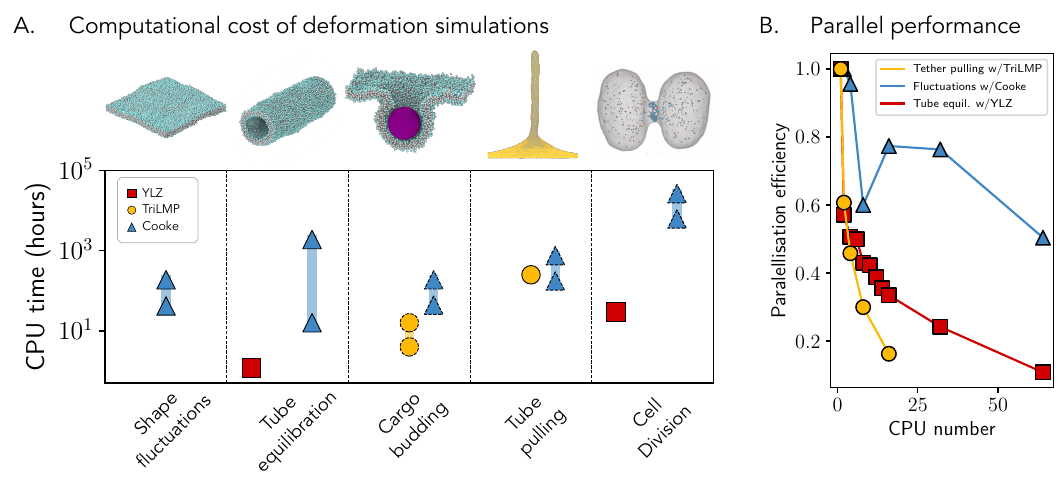}
    \caption{(A) CPU time (hours) spent in simulating different membrane deformations using the membrane models in Section~\ref{sec:tests}.
    For TriLMP, cargo budding was simulated using two vesicles with $N_{1} = 2562$ (with diameter $\sigma_1/\sigma \approx 30$) and $N_{2} = 10242$ ($\sigma_2/\sigma \approx 60$) vertices in the mesh; two cargo particles were tested with diameters $\sigma_{c,1}/\sigma_1 = 6$ and $\sigma_{c,2}/\sigma_2 = 6$, and the strength of the interaction was chosen to ensure sufficiently strong adhesion ($\varepsilon/k_B T = 10$). Dashed lines indicate that the model cannot accommodate the topological changes required to complete the deformation. The tube pulling results for TriLMP correspond to the time required to equilibrate and obtain the curves presented in Section~\ref{subsec:trilmp}.
    For the YLZ model, the CPU time for the tube equilibration was that for the simulations presented in Section~\ref{subsec:ylz}
    For the Cooke model, the CPU time ranges were obtained from simulation datasets appropriate for each measurement of shape fluctuations, tube equilibration and cargo budding; for this latter case we included the CPU time required for the equilibration of the final state.
    For tube pulling and cell division, we used estimates based on scaling the budding simulation by a factor of respectively 4 and 140, the latter corresponding to a cell of area $1\ \mu m^2$.
    (B) Comparison of the parallel performance of the three models. The drastic fall in efficiency for TriLMP is due to the partially parallel nature of the code: bond update moves, although computed in parallel~\cite{siggel2022} are currently serially implemented in TriLMP. Parallelisation in LAMMPS depends on how the simulation box is subdivided. The YLZ tube simulation was parallelised only by dividing the box along one axis, hence the monotonic curve. The Cooke fluctuation simulation was divided along two axes, which explains the dip at 8 CPUs.
    }
    \label{fig:simtime}
\end{figure*}

\subsection{Model comparison: on computational time}\label{subsec:timing}
The choice of a specific membrane model can ultimately be influenced by the computational resources available. As with any MD and MC simulation, computational time generally increases with the number of elements included in the model: for example, in an MD simulation, the more particles the system contains, the more equations of motion must be integrated, and thus the more expensive it becomes. Fig.~\ref{fig:simtime}A shows the different CPU timescales required to simulate and study typical membrane deformation phenomena such as fluctuations, endocytosis, tube extrusion or cell division. While the figure suggests a model like YLZ~\cite{Yuan2010} to be optimal for simulating membrane phenomena from the computational-time point of view, this is accompanied by a loss of resolution that the model entails compared to the Cooke model~\cite{Cooke2005}.

All three YLZ, TriLMP and Cooke run on code that is either fully or partially parallelised, which aids in reducing the total physical simulation time. However, it is important to know which parts of the computational model can be parallelised, and if so, how efficient this parallelisation is. The parallelisation efficiency is defined as $E = S/n$., i.e. the ratio between the speed-up of the code (that is, $S = T_1/T_n$, where $T_1$ is the run time in a single CPU core and $T_n$ is the runtime on $n$ CPU cores), and the number of cores $n$ used In Fig.~\ref{fig:simtime}B we compare the parallelisation efficiency of TriLMP, YLZ and Cooke models as a function of the number of CPUs.


\vspace{3mm}
\section{Discussion, perspective and open questions}

From cells to organelles to vesicles, the role of membranes as shielding and structuring barriers, and as a platform for protein encounters and assembly, is ubiquitous in biological systems. Therefore, membrane remodelling is fundamental for many cellular processes such as cell division, cell migration and endocytosis~\cite{frey2021}. Despite decades of remarkable progress in understanding these processes, numerous unsolved problems remain in the field of membrane biophysics. For example, understanding membrane bound protein self-organization and self-assembly processes, along with coupling these to non-equilibrium driving forces such as ATP/GTP hydrolysis, is the focus of much of today's biophysical research and reconstituted biology~\cite{Loose2008,Wu2016,dreher2020,baldauf2022}. By complementing novel experimental research in cell and membrane biophysics, computational membrane models have become a highly valuable tool to elucidate the key mechanisms at play in these systems.
However, entering the field of computational membrane physics can be challenging due to the multidisciplinary character of the subject. This pedagogical tutorial aims to lower the entry barrier by combining a state-of-the-art overview of the field with practical guidance to advise on how to pick and apply membrane models at the mesoscale.

Throughout the review, we have emphasized how mesoscale membrane models are built on different \textit{representations} (surface vs particle-based), cover different \textit{scales} (from single lipids up to whole cell membranes) and come with specific (dis)advantages.
We have shown that the choice of an appropriate membrane model depends significantly on the biophysical system under investigation. Indeed, by providing a hands-on tutorial for three distinct membrane models at the mesoscale, and comparing the three representations using criteria such as performance, adaptability and ability to represent the biophysical regime, we have demonstrated that choosing the 'best' model is often highly context-dependent. Nevertheless, one of the key points of our review is the notion that the process of finding a suitable model for a problem of interest can be systematised. Likewise, the breadth of models does not mean the reader must choose and stick to a single one for their project: we argue in favour of testing various membrane models under the same conditions to ensure that the results are independent of the used modelling techniques.

It is essential to know the limitations of the different models, and how to approach questions that lie precisely in the spaces which are difficult to model, as these are the spaces where new and exciting research takes place. Focusing on the limitations of a model sheds light on the gaps in knowledge in the field and can bring clarity to future perspectives and challenges.
As one of the future research directions, structural biologists and atomistic MD modellers have argued that the ultimate model for the membrane and for the full biological cell is the so-called \textit{digital twin}~\cite{beck2024}. Although such model is still very far from possible~\cite{deserno2009}, it is also unclear how much such a high resolution modelling alone can improve our understanding of the (emergent) mechanisms at play in biological systems. Moreover, in most cases, experimentally, we do not know the exact molecular composition of each individual cell, limiting building of such models even if they would be possible.

In our view, the intellectual value of coarse-grained mesoscale membrane models goes beyond their computational feasibility -- rather than being a substitute for computationally unfeasible models,  they represent an independent approach whose goal is to reveal key underlying physical principles, which are generalisable and go beyond chemical specificity. Precisely because top-down models deliberately simplify the processes under consideration, they enable us to build a physical intuition for the system, and to uncouple driving forces from one another. It is our opinion that coarse-grained mesoscale membrane models constitute a powerful, highly interpretable, readily available and relatively cheap tool that enables the study of biological matter at multiple scales, and that can be easily interfaced with the increasing amounts of experimental biological data and augmented by machine learning~\cite{Jung2023,Sahrmann2025}.

We predict that in the upcoming years the field of membrane biophysics will become ever more multi-disciplinary, more frequently coupled to experiments carried out in living cells as opposed to passive soft matter systems, often characterised by dynamical membrane heterogeneity and asymmetry, and inherently driven out-of-equilibrium. The constant back and forth between experimenters and modellers/simulators will be a norm, in particular due to the ever growing accessibility of quantitative cell biology experiments and high resolution imaging techniques~\cite{liu_super-resolution_2022,verweij_power_2021}.
Models will be needed to thoroughly map out and expand the experimentally accessible behaviour, as well as to predict and test physical mechanisms behind complex cell phenomena in order to guide future experiments.
Likewise, the expanding number of different modelling techniques will also trigger the need for more consistent mapping between the different models.
It is likely that models will incorporate various techniques at once, for instance by simultaneously combining several modelling approaches~\cite{beiter2024}, or including methods of artificial intelligence.
We hope that this guide will serve as a valuable reference point for both experimentalist and modellers in this exciting research space, so that models can be more easily learnt, compared and combined in the future.

\section*{Acknowledgements}
We thank Oded Farago, Angelo Cacciuto, Jeriann Beiter and Pietro Sillano for helpful discussions and a critical reading of the manuscript.
MMB and AP acknowledge funding by the European Union’s Horizon 2020 research and innovation programme under
Marie Skłodowska-Curie Grant Agreement No. 101034413.
FF acknowledges financial support by the NOMIS foundation.
BM and AŠ acknowledge funding by ERC Starting Grant “NEPA” 802960.
MA and AŠ acknowledge funding by the Volkswagen Foundation Grant Az 96727.

\section*{Author contributions}
MMB, FF, BM, MA, and AP contributed equally to this work.
All authors wrote and approved the manuscript.

\section*{Conflicts of interest}
There are no conflicts to declare.

\section*{Data availability}
Tutorial codes to reproduce the results in Section~\ref{sec:tests} are given as a repository~\cite{githubrepo}.





\bibliography{Bibliography} 
\bibliographystyle{rsc} 

\end{document}